\documentclass[a4]{revtex4}

\usepackage{graphics}

\newcommand{\angstrom}{\AA\ }
\newcommand{\rtilde}{\tilde{\mathbf{r}}}

\newcommand{\rcut}{\tilde{r}_{\text{cutoff}}}
\begin{document}

\title{The Role of Nucleobase Interactions in RNA Structure and Dynamics}

\author{%
Sandro Bottaro\,$^{1,*}$,
Francesco di Palma\,$^{1}$,
Giovanni Bussi \,$^{1,}$%
\footnote{To whom correspondence should be addressed.
Email: sbottaro@sissa.it, bussi@sissa.it
}}

\affiliation{%
$^{1}$
Scuola Internazionale Superiore di Studi Avanzati, International School for Advanced Studies, 
265, Via Bonomea I-34136 Trieste, Italy
}

\begin{abstract}
The intricate network of interactions observed in RNA three-dimensional structures is often described in terms of a multitude of 
geometrical properties, including helical parameters, base pairing/stacking, hydrogen bonding and backbone conformation.
We show that a simple molecular representation consisting in one oriented bead per nucleotide
can account for the fundamental structural properties of RNA. In this framework,
canonical Watson-Crick, non-Watson-Crick base-pairing and base-stacking interactions 
can be unambiguously identified within a well-defined interaction shell.
We validate this representation by performing two independent, complementary tests. 
First, we use it to construct a sequence-independent, knowledge-based scoring function for RNA structural prediction,
which compares favorably to fully atomistic, state-of-the-art techniques.
Second, we define a metric to measure deviation between RNA structures
that directly reports on the differences in the base-base interaction network. 
The effectiveness of this metric is tested with respect to the ability to discriminate between structurally and kinetically distant RNA conformations,
performing better compared to standard techniques. 
Taken together, our results suggest that this minimalist, nucleobase-centric representation
captures the main interactions that are relevant for describing RNA structure and dynamics.
\end{abstract}
\maketitle

\section{Introduction}
Ribonucleic acid (RNA) plays a fundamental role in a large variety of biological processes, such as enzymatic catalysis \cite{kruger1982self},
protein synthesis \cite{klein2004roles}, and gene regulation \cite{serganov2013decade}.  RNA molecules fold in a well-defined three-dimensional structure dictated by their nucleotide sequence \cite{tinoco1999rna}, that can be determined
by means of X-ray crystallography or nuclear magnetic resonance experiments \cite{tinoco1996nucleic}.
Despite recent important advances in the field, RNA structure determination is still a complex and expensive procedure.
Computational/theoretical models, and in particular structure prediction methods, ideally complement experiments, 
as they can in principle provide the tertiary, native structure for a given RNA sequence. 

Although traditionally considered simpler than the protein folding problem \cite{tinoco1999rna}, \emph{de novo} prediction of RNA three-dimensional structure still represents a challenging task \cite{cruz2012rna}. Atomistic molecular dynamics methods have been so far limited to the predictive folding of very small systems \cite{kührová2013computer,chen2013high}, whereas coarse-grained, physics-based models have proven useful in folding larger RNA structures \cite{ding2008ab,denesyuk2013coarse}. 
However, the most popular and successful RNA structure prediction methods are the so-called knowledge-based techniques, that typically employ fragment libraries for constructing RNA structures. These fragments are either combined using secondary structure information \cite{parisien2008mc,cao2011physics} or used to  generate a large number of plausible, putative structures that are subsequently ranked according to a scoring function \cite{das2007automated,frellsen2009probabilistic,das2010atomic,bernauer2011fully,capriotti2011all}. Knowledge-based methods often rely on two assumptions: First, that the main structural features of RNA molecules can be described in terms of few relevant observables. Second, that the  distribution of these observables in a dataset of experimental structures displays specific features, that can be used for generating and/or scoring three-dimensional structures
\cite{tanaka1976medium}. A crucial issue is the choice of the aforementioned observables.
While backbone atoms and torsional angles are the most natural choice for representing protein structures
\cite{ramachandran1963stereochemistry}, 
RNA molecules are more difficult to treat, due to the presence of a larger number of flexible backbone degrees of freedom 
and because of the important structural role of base-base interactions. For this reason RNA molecules are often described in terms of several structural observables, including backbone dihedrals \cite{murray2003rna,hershkovitz2006statistical}, pseudo dihedrals \cite{duarte1998stepping}, helical parameters \cite{saenger1984principles1}, hydrogen-bond networks, and stacking interactions \cite{leontis2001geometric,gendron2001quantitative}.

The proper choice of simplified, insightful descriptors for RNA molecules is not limited to the development of structure prediction methods,
but is a general and relevant issue in RNA structural analysis, as in the definition of metrics for comparing three-dimensional structures.
The well-known root mean square deviation (RMSD) after optimal superposition \cite{kabsch1976solution} is known to be suboptimal in the analysis of RNA structure, 
as it does not  provide reliable information about the differences in the base interaction network \cite{cruz2012rna,parisien2009new}.
\enlargethispage{-65.1pt}
\begin{figure*}[ht]
\begin{center}
\includegraphics{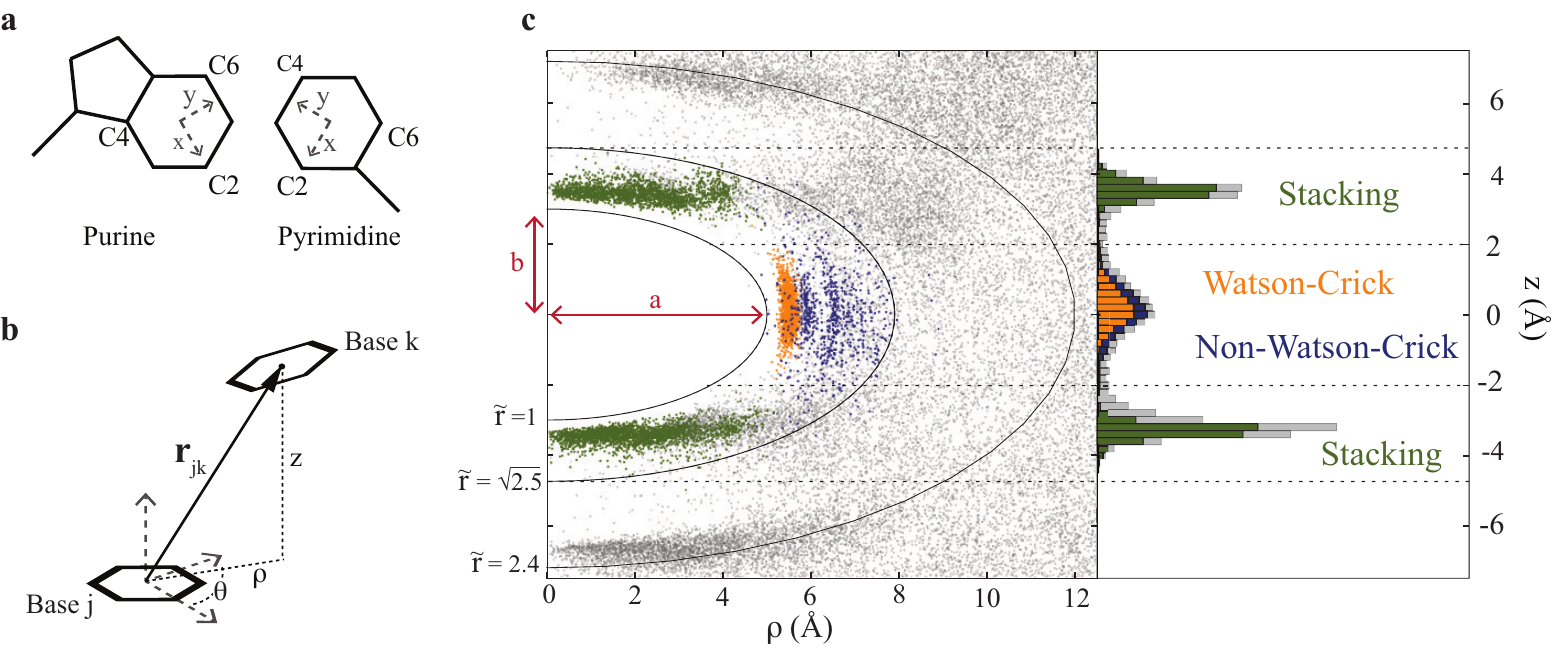}
\end{center}
\caption{(a) Definition of the local coordinate system for purines and pyrimidines.
(b) The vector $\mathbf{r}_{jk}$ describes the position of base ring $k$ in the coordinate system constructed on base ring $j$
and is expressed in cylindrical coordinates $\rho, \theta, z$. 
(c) Observed positions of neighboring bases in the crystal structure of the \textit{H. marismortui} large ribosomal subunit projected on the $z-\rho$ plane. Isolines of the ellipsoidal distance $\tilde{r}$ are shown.
Each point is colored according to the interaction type: canonical Watson-Crick pairs in orange,
base-base non Watson-Crick pairs in blue and stacking in green.  Base-sugar, base-phosphate and non-annotated pairs are shown in light grey.
On the right, empirical distribution along the $z$ coordinate obtained considering points with $\tilde{r}<\sqrt{2.5}$.
Pairing and stacking regions can be  unambiguously identified ($|z| \le 2.0$\angstrom and  $|z| > 2.0 $\AA, respectively).}
\label{fig:scheme1}
\end{figure*}
This motivated the development of alternative RNA-specific deviation measures based on selected geometrical properties.
These measures can be
used for comparing observed structures with {\em a priori} known patterns, as is done for instance in annotation methods \cite{yang2003tools} and in motif recognition algorithms \cite{sarver2008fr3d,apostolico2009finding,zhong2010rnamotifscan}. 
Other RNA specific similarity measures were introduced with the purpose of assessing the quality of RNA structural predictions \cite{cruz2012rna}
or for clustering/identifying recurrent structural motifs in large databases of known structures \cite{hershkovitz2006statistical,sarver2008fr3d}.
The definition of a proper structural deviation measure is also a central issue in the construction of Markov state models \cite{beauchamp2012simple},
that typically assume the geometric similarity to provide an adequate measure of kinetic proximity.

The analysis of RNA three-dimensional structure, the construction of a knowledge-based structure prediction method, and the 
definition of a structural deviation share a common trait: all of them are based on the apparently arbitrary
choice of the structural observables used to represent RNA.
Is there a simple representation of an RNA molecule that recapitulates its
key structural and dynamical properties?

In this Paper we introduce a minimalist representation for RNA molecules consisting in one oriented bead per nucleotide.
We first show that this description captures the fundamental base-pair and base-stacking interactions observed in experimental structures.
Using this representation we define a simple sequence-independent scoring function for RNA structure prediction, which
performs on par or better compared to existing atomistic techniques.
Based on the same molecular description, we then introduce a metric for calculating  structural deviation.
When compared with the standard RMSD measure, this metric correlates better with the physical time required to
explore the conformational space in molecular dynamics simulations. 
As a further proof of the capability of this metric to distinguish relevant structural differences, we show that it
can be successfully used for recognizing local structural motifs, reproducing
results obtained with state-of-the-art techniques. Finally, in the Discussion Section,
we summarize the analogies between these applications and
we proceed with an extensive comparison with existing techniques.
The software for performing the calculations is freely available at https://github.com/srnas/barnaba.

\section{Methods}
This Section is structured as follows: first, we introduce a simple representation for RNA three-dimensional structures
that has an intuitive interpretation in terms of base-stacking and base-pairing interactions.
We then use this molecular description to define i) a scoring function for RNA structure prediction ($\mathcal{E}$SCORE) and ii)
a metric for calculating deviations between RNA three dimensional structures ($\mathcal{E}$RMSD). 

\begin{figure*}[hb]
\begin{center}
\
\includegraphics{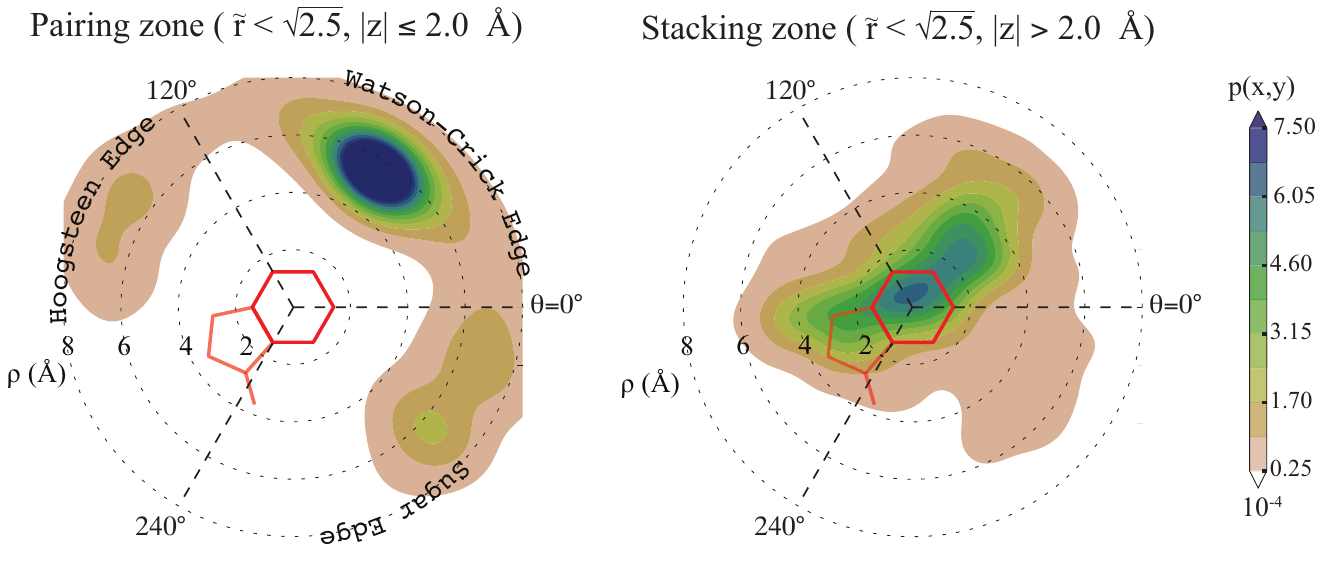}
\end{center}
\caption{Empirical density distributions of neighboring nucleobases ($\tilde{r}<\sqrt{2.5}$), obtained by projecting points belonging to the pairing and stacking zone of Fig.\ \ref{fig:scheme1} on the $\theta-\rho$ plane.
Densities are computed with a uniform binning in Cartesian coordinates and using a Gaussian kernel density estimation with bandwidth 0.25
\AA.
Hoogsteen, Watson-Crick, and Sugar edges  are indicated. The 6-membered and 5-membered (for purine only) rings are sketched in red.}
\label{fig:scheme2}
\end{figure*}

 \noindent \textbf{Molecular representation}
We construct a local coordinate system in the center of the 6-membered rings, as shown in Fig.\ \ref{fig:scheme1}a. 
Following this definition, the relative position and orientation between two nucleobases
is described by a vector $\mathbf{r}$, that is conveniently expressed in cylindrical coordinates
 $\rho$, $\theta$, and $z$ (Fig.\ \ref{fig:scheme1}b). Note that $\mathbf{r}$ is invariant for rotations around the axis connecting the 6-membered rings.
We highlight that this definition is similar to the local referentials introduced by Gendron and Major \cite{gendron2001quantitative}.
The use of a nucleotide-independent centroid makes it straightforward to compare and combine collection of position vectors deriving from different combinations
of nucleobases.  This is of particular importance for constructing the knowledge-based scoring function (see below).
The position vector $\mathbf{r}$ has an intuitive interpretation in terms of base-stacking and base-pairing interactions.
This aspect is illustrated in Fig.\ \ref{fig:scheme1}c, that shows the distribution of $\mathbf{r}$ vectors for all neighboring bases in the crystal structure of the \textit{Haloarcula marismortui} large ribosomal subunit (PDB code 1S72) \cite{klein2004roles}  projected on the $\rho$ and $z$ coordinates. In the figure,
different colors correspond to different types of interactions detected by MC-annotate.
Due to steric hindrance, no points are observed in a forbidden ellipsoidal region. Furthermore, 
almost all the base-stacking and base-pairing interactions ($\approx 99.6\%$) belong to a well-defined ellipsoidal shell.
It is therefore useful to introduce the anisotropic position vector
\begin{equation}\label{eq:rtilde}
\rtilde = \left( \frac{r_x}{a},\frac{r_y}{a} ,\frac{r_z}{b} \right) =  \left( \frac{\rho}{a}\cos{\theta},\frac{\rho}{a}\sin{\theta} ,\frac{z}{b} \right)
\end{equation}
with $a=5$\AA\ and $b=3$\AA, so that pairs of bases in the interaction shell are such that $1 < \tilde{r} < \sqrt{2.5}$.
The majority of base-base contacts lying in this interaction shell are annotated either 
as Watson-Crick/non-Watson-Crick or as base stacking, as detailed in Table \ref{tab:table1}.
Within this region we distinguish a \textit{pairing zone} and a \textit{stacking zone}, according to the type of featured interactions.
The tri-modal histogram in Fig.\ \ref{fig:scheme1}c shows that these two zones can be defined without ambiguity considering
pairs such that the projection of $\mathbf{r}$ along the $z$ axis is larger (stacking) or smaller (pairing) than 2 \AA.

\begin{table}
\caption{\textbf{Number of base-base interactions detected in the crystal structure of the \textit{H. marismortui} large ribosomal subunit.}} \label{tab:table1}
\begin{center}
\begin{small}
\begin{tabular}{ l c c} 
Interaction Type							&	Occurrences $^a$ 	& $\tilde{r}<\sqrt{2.5}^b$  \\
\hline
Stacking									&	 2328	& 2318 \\
Watson-Crick 								&	 723	        &  723  \\
Non-Watson-Crick (base-base)					&	 399		& 382 \\
Base-sugar/phosphate interactions	&	 781		& 398  \\
Not annotated 								& 	---		& 1214 	\\
\hline
\multicolumn3l{$^a$Number of interactions detected by MC-Annotate \cite{gendron2001quantitative}.}\\
\multicolumn3l{$^b$Base pair (j,k) is counted when both $\tilde{r}_{jk}$ and $\tilde{r}_{kj}<\sqrt{2.5}$.}
\end{tabular}
\end{small}
\end{center}
\end{table}

It is well known that the strength and nature of pairing and stacking interactions depend on the base-base distance, on the angle $\theta$ as well as on other angular parameters (e.g., twist, roll, tilt) in a non-trivial manner \cite{leontis2001geometric,vsponer2001electronic}. 
Such dependence can be observed in Fig.\ \ref{fig:scheme2}, where the points belonging to the pairing and stacking zone of Fig.\ \ref{fig:scheme1}c are projected on two separate $\rho$-$\theta$  planes. These distributions give an average picture containing contributions from different base pair types 
(purine-purine, purine-pyrimidine and pyrimidine-pyrimidine) and with weights dictated by the employed dataset. 
Nevertheless, the observations below hold also when considering the 16 possible combinations of base pairs individually and different datasets (see Supporting Data (SD) Figs.\ 1-3). 
In the pairing zone (Fig.\ \ref{fig:scheme2}, left panel) we first observe a dominant peak centered around  ($\rho$=$5.6$ \AA, $\theta$=$60^{\circ}$),
corresponding to the position of canonical Watson-Crick base pairs as well as wobble (GU) base pairs. 
The two other peaks correspond instead to base pairs interacting through the Hoogsteen or sugar edge \cite{das2007automated}.
One can also appreciate the absence of bases in the region occupied by the sugar ($190^{\circ}<\theta<290^{\circ}$).
The probability distribution in the stacking zone (Fig.\ \ref{fig:scheme2}, right panel) shows a broad peak in the proximity of the origin and extending up to
$\rho\approx4$\AA, which can be compared to the typical radius of the 6-membered ring ($\approx 1.4$\AA). 
This means that partial or negligible ring overlap is very frequent in RNA structures, as also 
observed in a seminal paper by Bugg \textit{et al.} \cite{bugg1971stereochemistry}.
This feature is more evident in pyrimidine-pyrimidine and purine-purine pairs, for which high overlap is the exception rather than the rule (see SD Fig.\ 3),
whereas overlap is systematically observed in  pyrimidine-purine pairs.
The fact that bases in the stacking zone are very often ``imbricated,'' similarly to roof tiles, rather than literally stacked
one on top of the other, does not imply that they are not interacting. 
Indeed, base-base interaction is not limited to $\pi$-$\pi$ stacking but also includes
electrostatic effects, London dispersion attraction, short range repulsion as well as 
backbone-induced effects \cite{vsponer2008nature}. 

\textbf{Scoring Function} The empirical distribution of all $\mathbf{r}$ vectors observed in experimental RNA structures displays very specific features, as described above.
We use these geometrical propensities to construct a knowledge based scoring function for RNA structure prediction. More precisely, we define a function of the atomic coordinates (in this case of the $\mathbf{r}$ vectors in a molecule) that quantifies the compatibility of a given RNA 3D conformation with respect to the expected distribution observed in native RNA structures. 
This scoring function, called $\mathcal{E}$SCORE, is defined as a weighted sum over all pairs of bases in a molecule:
\begin{equation}
\mathcal{E}\text{SCORE} = \sum_{j,k} p(\mathbf{r}_{jk})  \label{eq:score}
\end{equation}
Notice that both permutations of the $j,k$ indexes should be included in the sum since the vectors
$\mathbf{r}_{jk}$ and $\mathbf{r}_{kj}$ provide two partially independent pieces of information.
The weights $p(\mathbf{r})$ are given by the empirical probability distribution of nucleobases in the crystal structure of the \textit{H. marismortui} large ribosomal subunit. 
With this definition, a lower weight is assigned to structures with pairs of bases in relative positions not compatible with the training set, including those with steric clashes.
The probability distribution $p(\mathbf{r})$ is calculated considering only pairs of bases within the interaction shell ($\tilde{r} < \sqrt{2.5}$) and using a Gaussian kernel density estimation with bandwidth = $0.25$ \AA.  Note that $p(\mathbf{r})$ does not depend on the identity of the nucleotides, and the RNA molecule is therefore treated as a homopolymer.
The sum of the probabilities is used instead of their product, in order to decrease the impact of low-count regions
on the scoring function. This procedure has been first introduced in the field of hidden Markov modeling for speech recognition \cite{juang1997minimum}
and has a formal justification known as  ``summation hack'' \cite{minka}.
A non-redundant set of high resolution structures \cite{bernauer2011fully} was also employed as a training dataset for the $\mathcal{E}\text{SCORE}$, producing similar results. 
  
\textbf{Structural deviation} The collection of the scaled vectors $\rtilde$ (see Eq.~\ref{eq:rtilde}) in a RNA molecule
provides information about the relative base arrangement. One could thus
define a metric for measuring the distance between two RNA structures, $\alpha$ and $\beta$,  as
\begin{equation}
d= \sqrt{\frac{1}{N}\sum_{j,k} |\rtilde_{jk}^{\alpha} - \rtilde_{jk}^{\beta}|^2}
\label{eq:distance0}
\end{equation}
Similarly to Eq. \ref{eq:score}, also here both permutations of the $j,k$ indexes should be included in the sum.
The $d$ quantity is highly non-local, because large differences in distant pairs give large contributions to the sum.
To remove this effect it is convenient to introduce a cutoff distance $\rcut$ in the same spirit as we did for the scoring function
discussed above. However, we notice that a cutoff would introduce discontinuities in the metric.
A possible solution is to introduce a function that maps the vectors $\rtilde$ to new vectors $\mathbf{G}(\rtilde)$ with the following properties:
\begin{enumerate}
\item $|\mathbf{G}(\rtilde^{\alpha}) - \mathbf{G}(\rtilde^{\beta})| \approx |\rtilde^{\alpha} - \rtilde^{\beta}| $ 
if $\tilde{r}^{\alpha},\tilde{r}^{\beta} \ll \rcut$.
\item  $|\mathbf{G}(\rtilde^{\alpha}) - \mathbf{G}(\rtilde^{\beta})|  = 0  $ if $\tilde{r}^{\alpha},\tilde{r}^{\beta} \ge \rcut $.
\item $\mathbf{G}(\rtilde)$ is a continuous function.
\end{enumerate}
There is no requirement here on the dimensionality of the $\mathbf{G}$ vector.
The above properties are satisfied by the 4-dimensional vectorial function 
\begin{equation}
   \mathbf{G}(\rtilde) = \left(
      \begin{array}{c}
	\sin{(\gamma \tilde{r}})\frac{\tilde{r}_x}{\tilde{r}} \\
	\sin{(\gamma \tilde{r}}) \frac{\tilde{r}_y}{\tilde{r}} \\
	\sin{(\gamma \tilde{r}}) \frac{\tilde{r}_z}{\tilde{r}} \\
 	1+\cos{(\gamma \tilde{r})}
\end{array} \right) \times \frac{\Theta(\rcut - \tilde{r})}{\gamma}
\label{eq:g}
 \end{equation}
where $\gamma = \pi/\rcut$ and $\Theta$ is the Heaviside step function.
We note that the dependence of $\mathbf{G}$ on the direction of $\rtilde$
vanishes as $\tilde{r}$ approaches the cutoff value $\rcut$.
We thus set 
$\rcut=2.4$, so that for the typical distances between stacked and paired bases
($\tilde{r} < \sqrt{2.5}$, see Fig.\ \ref{fig:scheme1}c) the contribution of the $\rtilde$ directionality is significant.
In-depth discussion on Eq.~\ref{eq:g} can be found in SD Text.~4.
Having defined the mapping function $\mathbf{G}(\rtilde)$,  the $\mathcal{E}\text{RMSD}$ distance reads
\begin{equation}
\mathcal{E}\text{RMSD} = \sqrt{\frac{1}{N}\sum_{j, k} |\mathbf{G}(\rtilde_{jk}^{\alpha}) - \mathbf{G}(\rtilde_{jk}^{\beta})|^2}
\label{eq:distance}
\end{equation}
Note that $\mathcal{E}\text{RMSD}$ is proportional to the Euclidean distance between $\mathbf{G}$ vectors, which are non-linear functions of the atomic
coordinates. As such, $\mathcal{E}\text{RMSD}$ is positive semidefinite, symmetric, and satisfies triangular inequality.
During the developmental stage we tested a non-vectorial version of the $\mathcal{E}$RMSD:
instead of using the vectorial function $\mathbf{G}(\rtilde)$, we considered the scalar, continuous function 
\begin{equation}
\label{eq:g-s}
G_s(\rtilde) = (\rcut - \tilde{r}) \times \Theta(\tilde{r}_{\text{cutoff}}-\tilde{r})
\end{equation}
which is similar to a contact-map distance \cite{bonomi2008unfolded,cossio2011similarity}.
This scalar version differs from $\mathcal{E}\text{RMSD}$ when considering 
structures with 4 nucleotides or less, while the two quantities are highly correlated when analyzing larger
structures (Figure SD 5).

\section{Results}
 
\subsection{Scoring Function for RNA Structure Prediction}
We assess the quality of the $\mathcal{E}$SCORE on its capability of recognizing the folded state among a set of decoys \cite{simons1997assembly}.
We consider here a total of 39 different decoy sets previously used to benchmark other knowledge-based scoring functions:
20 decoy sets generated \textit{de novo} using the FARNA algorithm \cite{das2007automated}  and 19 additional decoy sets from the work of Bernauer \textit{et al.} \cite{bernauer2011fully} obtained by gradual deformation of the native structures.
As a measure of performance we use the normalized rank \cite{cossio2012simple}, 
defined as the percentage of decoys scoring better than the native structure. 
In order to simulate a realistic structure prediction experiment, we additionally report the RMSD from native of the best scoring structure in the decoy set.
 Because the RMSD has been reported to be a suboptimal indicator of structural proximity for RNA,
 we also compute the interaction fidelity network (INF), which measures the overlap between annotations in two structures \cite{parisien2009new}, as well as the $\mathcal{E}$RMSD introduced here.
Equivalent analyses are performed using two state-of-the-art scoring functions, namely the FARFAR  \cite{das2010atomic} and the all-atom knowledge-based scoring function RASP \cite{capriotti2011all}.
Both FARFAR and RASP are defined at atomistic resolution. FARFAR combines the low resolution FARNA potential with explicit terms for solvation and hydrogen bonding, while RASP is a statistical potential based on pairwise interatomic distances.
These two scoring functions are among the best available \cite{cruz2012rna} and the only ones for which it was possible to perform automatic scoring of a large set of decoys.

\begin{small}
 \begin{figure}[h]
 \begin{center}
\includegraphics{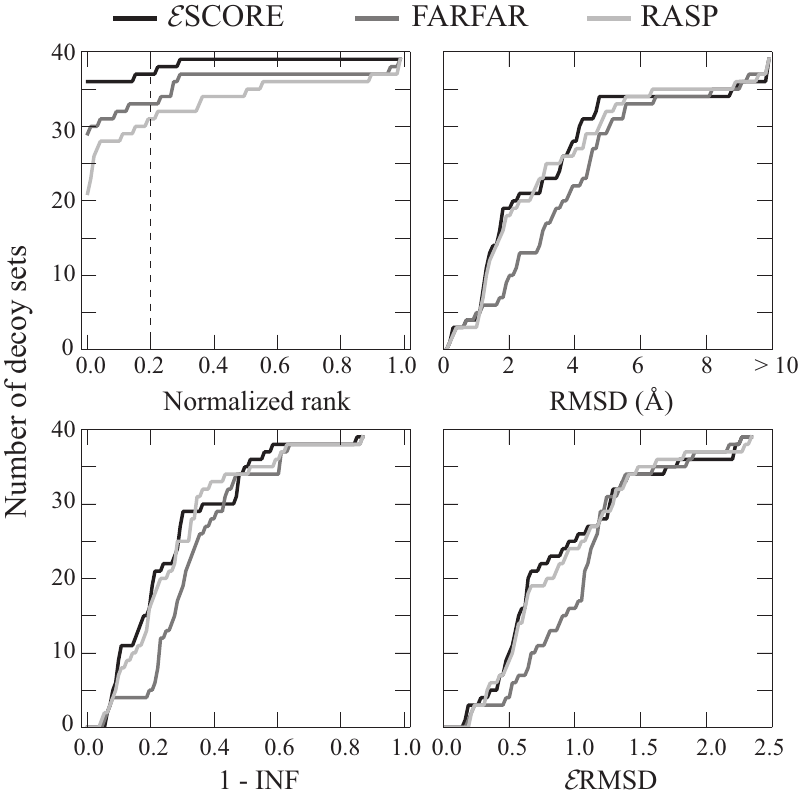}
\end{center}
\caption{ Quantitative comparison of our scoring function $\mathcal{E}$SCORE with FARFAR \cite{das2010atomic} and RASP \cite{capriotti2011all}
on 39 different decoy sets. In the upper left panel, number of decoy sets with normalized rank lower than the value on the horizontal axis. 
For example, the normalized rank is $0.2$ or less for 37 out of 39 decoy sets using the $\mathcal{E}$SCORE,  33/39 using FARFAR, and 31/39 using RASP.
The faster the curve reaches 39, the better the performance of the scoring function. In the other panels,
equivalent plots for RMSD, INF \cite{parisien2009new}, and $\mathcal{E}$RMSD of the best scoring decoy to the native structure, as labeled.
See Table SD6 for a complete list of results.} 
\label{fig:ranks}
\end{figure}
\end{small}
Figure \ref{fig:ranks} summarizes the results obtained on all the 39 decoy sets. For each method we report the
number of decoy sets below the value indicated on the horizontal axis. Thus, the better is a scoring function, the faster the curve reaches the maximum number
39.
In general, whereas FARFAR performs well in ranking and 
RASP in finding the best decoys, $\mathcal{E}$SCORE performs well in both tasks (See also Table SD 6). This is a remarkable result considering that $\mathcal{E}$SCORE employs a coarse-grained representation (one bead per nucleotide) and is not sequence dependent.
The scoring results on all RNA decoys, together with a short discussion of the cases for which  $\mathcal{E}$SCORE fails to identify correctly the native structure can be found in SD Figs.\ 7-8.

\subsection{RNA Structural Deviation}
The definition of an accurate measure of structural deviation for comparing RNA three-dimensional structures is a relevant problem,
as the standard RMSD approach does not report on differences of base-stacking and base-pairing patterns, 
that are the fundamental, key features of RNA molecules \cite{gendron2001quantitative}.
As a striking example, the RMSD between an ideal A-form double helix $^{\texttt{GGGG}}_{\texttt{CCCC}}$ and the mismatched structure $^{\texttt{\phantom{G}GGGG}}_{\texttt{CCCC \phantom{C}}}$
is 1.9\AA, meaning that the threshold of 4\angstrom that is sometimes employed to consider two structures as similar \cite{chen2013high} may not be sufficiently strict.
A similar issue arises in kinetic modeling of proteins, where the assumption that geometrically close configurations (in terms of the RMSD distance) are also kinetically close has been shown to be problematic in discretizing the register shift dynamics of $\beta$ strands \cite{beauchamp2012simple}.
\begin{small}
\begin{figure}[b]
\begin{center}
\includegraphics{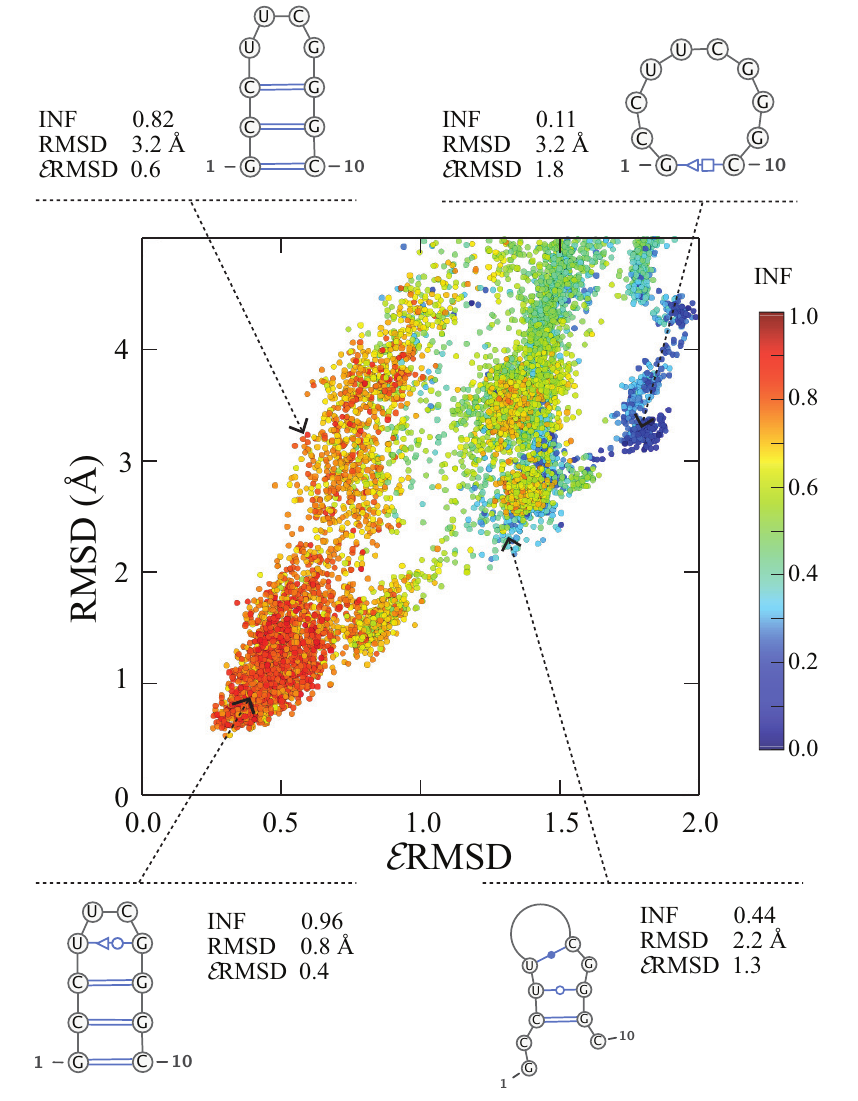}
\end{center}
\caption{Heavy-atom RMSD from native versus $\mathcal{E}$RMSD of snapshots taken from an atomistic molecular dynamics simulation.  
The difference in the secondary structure between the native state and samples is calculated using the INF measure, which ranges from 1 (identical secondary structure) to zero (completely different). Very different secondary structures can be found below 4 \angstrom RMSD. Four selected structures with different values of RMSD, $\mathcal{E}$RMSD and INF are shown, together with their secondary structure. Figures were generated with Varna \cite{darty2009varna}.}
\label{fig:scheme4}
\end{figure}
\end{small}

In order to elucidate the difference between the standard RMSD and the $\mathcal{E}$RMSD, we compute these two quantities on a series of snapshots taken from an atomistic molecular dynamics simulation of a short stem-loop. These structures have been obtained by repeatedly melting and folding (see SD Text 9), and thus include near-native structures, partially folded, misfolded as well as extended conformations. Figure\ 4  shows that structures with similar RMSD do not necessarily present the same base-base interaction network of the native state: in fact, the secondary structure can be substantially different.
Conversely, the $\mathcal{E}$RMSD discriminates structures with near-native base-base contacts
($\mathcal{E}$RMSD $< 0.8$) from structures with non-native base-base interactions ($\mathcal{E}$RMSD$>1$). Note that in general low $\mathcal{E}$RMSD correspond to high INF values, except for the region around $\mathcal{E}$RMSD$=1.5$, RMSD$=2.7$. In this region we observe structures with formed stem and distorted loop (high INF) but also structures with 
formed loop and incorrect stem (low INF). 
While Fig.\ 4 serves as an illustrative example, the validity of the $\mathcal{E}$RMSD is more rigorously assessed by performing two tests that
probe the accuracy of the method in the analysis of structural and dynamical properties of RNA molecules.
More precisely, in the following two Subsections we demonstrate  i) that the $\mathcal{E}$RMSD is a good measure of kinetic proximity between configurations, performing considerably better 
compared to standard structural distances and ii) that the $\mathcal{E}$RMSD is highly specific in recognizing  known structural motifs within a large set of crystallographic structures.\\
\textbf{Structural and Kinetic Proximity}
The correspondence between a structural deviation (RMSD, $\mathcal{E}$RMSD, etc.) and 
the kinetic distance (defined as the time needed to evolve from one configuration to the other),
can be quantitatively assessed by computing the coefficient of variation (CV).

\begin{figure}[b]
\begin{center}
\includegraphics{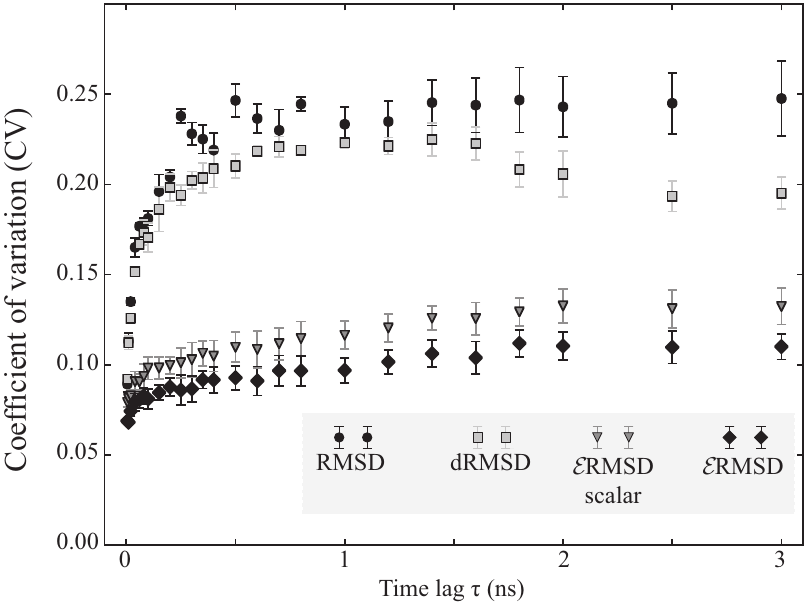}
\end{center}
\caption{Coefficient of variation calculated on a 100ns molecular dynamics simulation of the \textit{add} riboswitch. The coefficient of variation for the $\mathcal{E}$RMSD is compared with heavy-atoms RMSD, with the distance RMSD (dRMSD) and with the scalar variant of the $\mathcal{E}$RMSD. Standard deviations are calculated using a blocking procedure.}
\label{fig:scheme5}
\end{figure} 

The CV of a structural deviation $d$ is defined as the ratio of the standard deviation to the mean
calculated on structures separated by a time lag $\tau$
\begin{equation}
CV_d(\tau) = \frac{\sigma_d(\tau)}{\langle d(\tau)\rangle}
\end{equation}
To a large CV corresponds a large standard deviation, and therefore a high uncertainty between geometric and kinetic distance \cite{zhou2012distribution}.
Note that the standard deviation is normalized with the mean, and the CV is thus scale-invariant. 
In the approximation that the deviations at fixed time lag are normally distributed, this analysis is equivalent to
the histogram overlap discussed in Ref.\ \cite{cossio2011similarity}.
The connection between kinetic and structural distance is inevitably limited by the fact that
the typical time necessary for a transition is in general different from
the time necessary for the reversed transition, in contrast with the symmetric definition of structural
deviations. Nevertheless, this analysis provides insightful and intuitive information
for the purpose of qualitatively comparing different metrics.

In Fig.\ 5 we show the CV as a function of time separation in 
molecular dynamics simulation of the \textit{add} riboswitch  \cite{serganov2004structural,di2013ligand}.
The molecular dynamics simulation was initialized in the native state and remained stable throughout the 100ns run, 
while experiencing non trivial dynamics such as base-pairing breakage/formation and double-helix breathing. 
The CV for $\mathcal{E}$RMSD is compared with RMSD and with dRMSD.
Additionally, the scalar variant of $\mathcal{E}$RMSD, based on Eq.~\ref{eq:g-s}, is also presented.

We first observe that for short time separations the correspondence between geometric and kinetic distance is high (low CV) in all cases, and it diminishes at large time lag. 
The CV for the $\mathcal{E}$RMSD is consistently lower compared to both RMSD and dRMSD, suggesting this measure to better reflect the temporal separation between configurations.  
We also note that, in this test, the scalar and vectorial version of the  $\mathcal{E}$RMSD behave very similarly.\\
\textbf{Search of Structural Motifs} 
 \begin{table}[b]
 \caption{Structural search of known motifs within the RNA motif atlas. 
The table summarizes the number of found motifs compared with FR3D (http://rna.bgsu.edu/rna3dhub/motifs). See SD Text 11 for a full list.}\label{tab:table3}
\begin{center}
 \begin{small}
\begin{tabular}{ cc c c}
Motif 	&  Reference PDB & Found motifs  & False positive \\
\hline 
T-LOOP 	& 3RG5	& 57/66	         & 17  \\	%
GNRA       & 4A1B	& 217/219 	 & 63 \\	%
C-LOOP    & 3V2F	& 18 /22   	& 7  \\	%
D-SHEAR & 1SAQ	& 23/26	  	& 3  \\	%
T-SHEAR  & 3V2F	& 24/31     	& 18 \\	%
K-TURN	&  3GX5	&	10/22       	&  15 \\	%
SARCIN	&  1Q96	&	13/16	  &  3 	%
\end{tabular}
\end{small}
\end{center}
\end{table}
RNA molecules display a great variety of base-base interaction patterns (RNA motifs) that play a key role 
in RNA folding and that mediate important biological processes \cite{moore1999structural}. A structural 
deviation should therefore accurately capture such RNA-specific features.
This property can be assessed by searching for known RNA structural motifs
in the database of experimentally solved structures.
 Here, we use the $\mathcal{E}$RMSD to search for 5 known internal loop motifs (K-Turn, C-loop, Sarcin Ricin, triple sheared, double sheared) and two hairpin loops (T-loop and GNRA) within the RNA structure atlas database \cite{petrov2013automated} release 1.43, composed by 772 crystal structures. 
As a reference, we compared the results with  ``find RNA 3D''  (FR3D) \cite{sarver2008fr3d}, the most extensively used and carefully tested method for RNA motif recognition. 
FR3D employs a reduced representation with one centroid per nucleobase, and computes
the geometrical deviation by combining fitting and orientation errors after optimal superposition.
Reference and motif structures were obtained from the RNA 3D motif atlas release 1.13. 
Except for the K-turn motif, we were able to correctly identify almost all FR3D matches, including motifs with bulged bases (Table \ref{tab:table3}). 
By visual inspection, it can be seen that most of the false positive (i.e., motifs that were found using the $\mathcal{E}$RMSD and not by FR3D) are either  \textit{bona fide} matches or known motif variants, while the residual discrepancies are due to differences in the employed cutoff value \cite{petrov2013automated}.  
We refer the reader to SD Text 10 for a detailed description of the search strategy and to SD Text 11 for the full list of results.

\section{Discussion}
A fully atomistic description of an RNA molecule requires on average $\approx20$ non-hydrogen atoms per nucleotide,
and thus $\approx60$ degrees of freedom per nucleotide.
Biologically relevant RNA structures are composed of hundreds or even thousands of nucleotides that display an intricate network of interactions. 
 The complexity of the problem is further increased when considering that many RNA molecules are highly dynamic entities, and multiple chain configurations are needed for describing their mechanism and function \cite{serganov2013decade}.  
In order to provide a simpler representation, RNA structures are usually analyzed in terms of sugar pucker conformations, backbone dihedrals \cite{murray2003rna,hershkovitz2006statistical} or pseudo dihedrals \cite{duarte1998stepping}.
Complementary approaches consider instead base-pair parameters (propeller, slide, roll, twist, etc.) and hydrogen bonding  to classify base-pair interactions in terms of recurrent geometrical configurations observed in experimentally solved RNA structures \cite{saenger1984principles1,leontis2001geometric,gendron2001quantitative}.

A key result of the present work is that the main structural and dynamical properties of RNA  can be described in terms of the
spatial arrangement of the nucleobases only.  Since nucleobases are substantially rigid, this implies $6$ degrees of freedom per nucleotide.
This observation is in full accordance with previous analysis of RNA crystal structures  \cite{gendron2001quantitative,lemieux2006automated},
where the position and orientation of nucleobases only were considered. This work complements the analysis by showing that nucleobase position and orientation
are sufficient to correctly describe the RNA dynamics as obtained from explicit solvent molecular dynamics simulations.

As shown in Fig.\ \ref{fig:scheme1}, RNA base-base interactions mainly  occur in a restricted neighborhood of the nucleobase that is well approximated by an ellipsoidal shell. Since specific regions of the ellipsoidal shell surrounding nucleobases correspond to different types of pairing and stacking interactions, it is in principle possible to infer secondary structure information directly from the pairing and stacking plots of Fig.\ \ref{fig:scheme2}. This procedure is formally similar to the one employed in the analysis of protein structures, where the angular preferences of the backbone approximately correspond to different secondary structures \cite{ramachandran1963stereochemistry}.  Our approach depends on nucleobase positions only and is thus complementary to that proposed in Ref. \cite{duarte1998stepping}, where backbone pseudo dihedrals were used to classify RNA secondary structures.

By considering only the relative spatial arrangement of nucleobases, we introduce a knowledge-based scoring function for RNA structure prediction ($\mathcal{E}$SCORE). In contrast with several studies underlining the importance of an atomistic description for an accurate scoring  \cite{das2010atomic,bernauer2011fully,capriotti2011all}, our work shows that a minimalist description of
 nucleobase arrangement is sufficient to produce equivalent or better results.  
 Two aspects are worth highlighting. First, while the backbone can be ignored for scoring, 
 one cannot avoid including the chain connectivity in the sampling algorithm used to generate the decoys.    
Second, the outcome of the scoring benchmarks is strongly dependent on the employed decoy sets. 
Some of the decoy sets generated using the FARNA algorithm are very challenging, as 
all scoring functions ($\mathcal{E}$SCORE, FARFAR and RASP) are highly degenerate, and assign good scores to structures that are very far from native 
(see e.g.\ SD Fig. 6, decoy set 1A4D and Ref.\ \cite{bernauer2011fully}).
These decoys can be therefore used to improve the current methodologies for \textit{de novo} predictions of RNA structure.

A second aspect intimately connected with the choice of molecular representation is the definition of a proper structural proximity measure.
Typically, structural deviations are used either to analyze and compare three-dimensional structures (e.g., RMSD, dRMSD, INF  \cite{parisien2009new}) or to 
recognize and cluster RNA structural motifs \cite{apostolico2009finding,sarver2008fr3d,zhong2010rnamotifscan}.
The $\mathcal{E}$RMSD introduced here is suitable for both tasks. 
A distinctive feature of the $\mathcal{E}$RMSD is the use of the ellipsoidal metric $\tilde{r}$,
that naturally stems from the empirical data and that takes into account both relative distance and orientation of nucleobases.
Interestingly, the intrinsic base anisotropy has been also discussed in the context of the electronic polarizability of nucleobases \cite{vsponer2008nature} and anisotropic interactions have been used in coarse-grain models for proteins \cite{fogolari1996modeling}.
We note that the $\mathcal{E}$RMSD is conceptually reminiscent of the INF measure \cite{parisien2009new}, that also considers the differences in the network of base-pairing and base stacking interactions.
With respect to INF, however, the $\mathcal{E}$RMSD presents important advantages: i) it does not depend on the annotation scheme employed and therefore on the quality of the structure, ii) it is a continuous function of the atomic coordinates, and iii)  it is well defined even when the annotation is not informative (e.g., for unfolded states, unstructured, or single-stranded RNAs).
The last two items make the $\mathcal{E}$RMSD particularly well-suited for studying dynamical processes, such as RNA folding or binding events. 
The $\mathcal{E}$RMSD also shares a number of similarities with the structural distance used in FR3D for recognizing three-dimensional motifs \cite{sarver2008fr3d}.  However, in the $\mathcal{E}$RMSD calculation we introduce 
important simplifications: i) the definition of the local coordinate system is not nucleobase-specific and requires the knowledge of three atoms per nucleobase only (Fig.\ \ref{fig:scheme1}a),
ii) the $\mathcal{E}$RMSD calculation does not require least square fitting, and iii) the problem of combining the positional and orientational base-base distance is circumvented by introducing an ellipsoidal metric.
The fact that the two methods produce similar results (Table 2) strongly suggests that for most practical uses the $\mathcal{E}$RMSD is accurate enough to capture the specific pattern of base-base interactions.

In this study we introduce a minimalist representation of RNA three-dimensional structures that solely considers 
the relative orientation and position of nucleobases.
Based on this representation, we construct a scoring function for RNA structure prediction and we define
a metric for calculating deviation between RNA three-dimensional structures.
In both cases, the results are on par or better compared with state-of-the-art techniques.
Taken together, the presented results suggest that this minimalist representation captures the main interactions that are relevant for describing RNA structure and dynamics.
As a final remark, we note that the methodology presented here can be readily applied to DNA molecules as well.

\section{Availability} 
Python scripts to compute $\mathcal{E}$SCORE and $\mathcal{E}$RMSD
from PDB structures are available at https://github.com/srnas/barnaba.

\section{Supplementary Data}
Supplementary Data are available at NAR Online.
Supplementary References \cite{jorgensen1983comparison,darden1993particle,hess1997lincs,berendsen1984molecular,ennifar2000crystal,hornak2006comparison,perez2007refinement,bussi2007canonical,banas2010performance,pronk2013gromacs,tribello2014plumed}.

\section{Funding}
The research leading to these results has received funding from the European Research Council under the European Union's Seventh Framework Programme (FP/2007-2013) / ERC Grant Agreement n. 306662, S-RNA-S.

\subsubsection{Conflict of interest statement.} None declared.

\section{Acknowledgments}
We are grateful to Francesco Colizzi, Richard Andre Cunha, Eli Hershkovits, Alessandro Laio, Gian Gaetano Tartaglia, Eric Westhof, and Craig Zirbel for
carefully reading the manuscript and providing several enlightening suggestions. Massimiliano Bonomi, Thomas Hamelryck and Andrea Pagnani are also acknowledged for useful discussions.

\bibliographystyle{nar}

\end{document}